\documentclass[aps,groupedaddress,showpacs,epsfig,floats,twocolumn]{revtex4}
\usepackage{graphicx}
\begin{document}

\title{Magnetic phase shifter for superconducting qubit}
\author{D. S. Golubovi\'{c}, W. V. Pogosov, M. Morelle and V. V. Moshchalkov}
\affiliation{Nanoscale Superconductivity and Magnetism Group,
Laboratory for Solid State Physics and Magnetism, K. U. Leuven,
Celestijnenlaan 200 D, B-3001 Leuven, Belgium}

\begin{abstract}
We have designed and investigated a contactless magnetic phase
shifter for flux-based superconducting qubits. The phase shifter
is realized by placing a perpendicularly magnetized dot at the
centre of a superconducting loop. The flux generated by this
magnetic dot gives rise to an additional shielding current in the
loop, which, in turn, induces a phase shift. By modifying the
parameters of the dot an arbitrary phase shift can be generated in
the loop. This magnetic phase shifter can, therefore, be used as
an external current source in superconducting circuits, as well as
a suitable tool to study fractional Josephson vortices.
\end{abstract}
\pacs{03.67.Lx, 74.78.Na., 75.75.+a} \maketitle

Various macroscopic solid state structures can exhibit quantum
behaviour, potentially interesting for quantum computing
\cite{mon}. A typical example is a superconducting flux qubit
based on Josephson junctions, which operates by utilizing the
degeneracy of the two equal and opposite persistent currents at
half-integer flux $(n+1/2)\Phi _{0}$ through the superconducting
circuit, where $\Phi _{0}$ is the superconducting flux quantum
(\cite{qb} and references therein). The degeneracy is lifted by
the charging energy and the two distinct quantum states
$|0\rangle$ and $|1\rangle$ are associated with the opposite
circulation of the superconducting condensate. A resonant external
excitation can make the superconducting condensate oscillate
coherently between these two states \cite{qb}. Although an
external applied field has been used in the experiments to
generate the flux necessary for $\pi$-shift in the superconducting
qubit, it is thought that the fluctuations of the external flux
(flux noise) present a major source of decoherence \cite{nob}.
Therefore, it has been a challenge to incorporate $\pi$-shift in
the qubit and enable its operation without an external bias field.
A number of structures using high-$T_{c}$ superconductors have
been proposed (\cite{ht} and references therein). In high-$T_{c}$
superconductors, due to the predominant $d_{x^{2}-y^{2}}$ symmetry
of the order parameter, $\pi$-shift can inherently be gained if,
for example, the interfaces of the Josephson contacts are made of
two superconductors rotated by $\pi/2$ in the {\it ab}-plane.
However, this particular symmetry poses a fundamental constraint
on the overall coherence of the high-$T_{c}$ superconducting
qubits. Since the superconducting gap $\Delta $ is zero along the
nodal directions, normal quasiparticles that are inherently
incoherent, are present even at very low temperatures. For this
reason, the application of high-$T_{c}$ superconductors is a
trade-off in terms of coherence: decoherence due to the flux noise
is eliminated, but an additional source of decoherence related to
the presence of normal quasiparticles is introduced.

In this Letter we propose a new practical realization of a phase
shift for the superconducting qubit. The phase shift is achieved
by placing a magnetic dot with the perpendicular magnetization at
the centre of a loop made of a {\it conventional} $s$-wave
superconductor. The flux generated by the dot creates an
additional current in the superconducting loop giving rise to a
phase shift. We believe that the proposed design has several
advantages. First and foremost, the phase shift is a result of a
quite basic and general property of superconductors and can be
implemented without any limitations. It does not require $d$-wave
symmetry of the order parameter, nor does it put any constraints
on the interfaces of Josephson junctions, as with high-$T_{c}$
\begin{figure}[hbt] \centering
\includegraphics*[width=8.5cm]{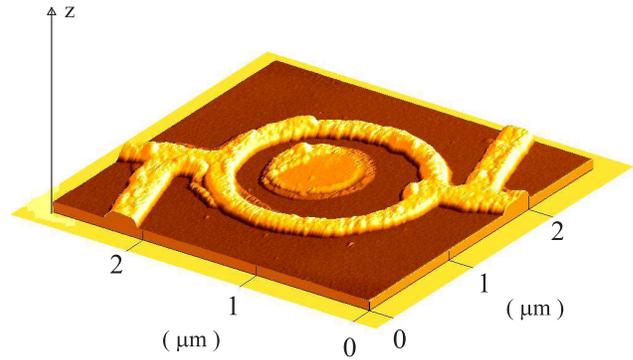}
\caption{An AFM image of the sample B. \label{afm}}
\end{figure}
superconductors. More importantly, the phase shift is achieved
with an $s$-wave superconductor, and, therefore, no additional
decoherence appears in the system, as in the case of the $d$-wave
superconductors. As magnetic dot is separated from the
superconducting loop it cannot, by all means, adversely affect the
operation of the qubit. The generated flux, and consequently, the
persistent current in the loop are stable. Technologically, the
fabrication procedures for conventional nanostructured
superconductors and ferromagnets have been mastered and can be
carried out routinely. By conveniently varying the parameters of
the dot it is possible to introduce {\it any} phase shift in the
loop. The magnetic phase shifter can, therefore, be used as an
external current source with a high stability. Furthermore, it may
well be applied for phase biasing in the experimental study of
fractional Josephson vortices \cite{gold}. Since the phase shifter
can be used with $s$-wave superconducting circuits, the
integrability and scalability of the qubit are not deteriorated.
\begin{figure}[hbt]
\centering
\includegraphics*[width=8.5cm]{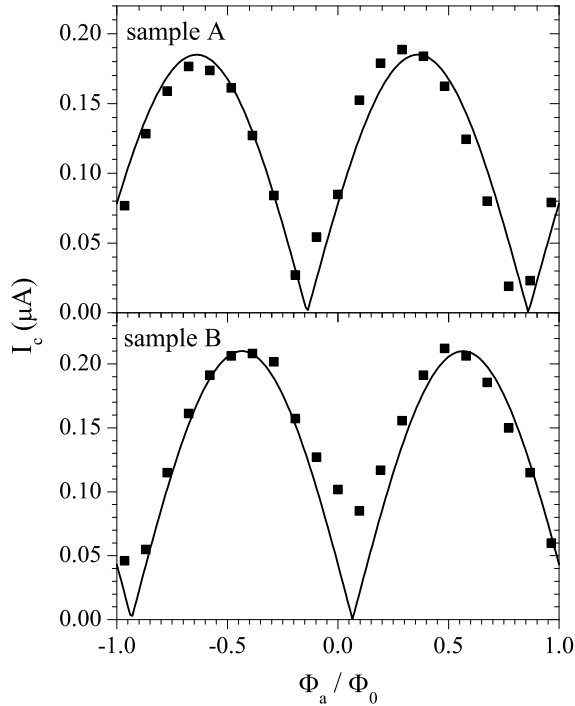}
\caption{The critical current of the structures versus the
normalized applied flux.  The measurements were taken with the
field step of $0.1\Phi_{a} /\Phi _{0}$ at $99.5\,\%$ of the
maximum critical temperature $T_{cm}=1.3105\,$K. The solid line is
the theoretical curve (Eq. (\ref{squid})) \label{icb}}
\end{figure}

The sample was prepared with electron beam lithography in three
phases. For the details on the fabrication procedure we refer to
Ref. \cite{ja,japrb}. The superconducting loop is made up of
$46\,$nm thick Al, with the inner and outer radii $680\,$nm and
$915\,$nm, respectively, whereas the radii of magnetic dots are
$174\,$nm (hereafter 'sample A') and $350\,$nm (hereafter 'sample
B'), respectively. The dots consist of 10 bilayers of $0.4\,$nm Co
and $1\,$nm Pd, with a $2.5\,$nm Pd buffer layer. This composition
of the dots has been chosen since it provides a sufficiently high
coercive field of approximately $150\,mT$, as well as a complete
remanence and nearly perfectly rectangular shape of the hysteresis
loop. Prior to the measurements the dot was saturated in the field
of $800\,$mT. The external magnetic field was being varied within
the range $-30\leq B_{a}\leq 30\,$[mT], so that the magnetization
of the dot remained unaltered during the measurements. The mean
free path of Al, estimated from the resistance of the
co-evaporated Al sample at $4.2\,$K, is $l\approx 16\,$nm, whereas
the Ginzburg-Landau coherence length is $\xi (0) \approx 138\,$nm.
Fig. \ref{afm} shows an atomic force micrograph of the sample B.
The surface of the Al loop, as well as a part of magnetic dot seem
to be corrugated. This is just due to the presence of the remains
of the electron beam resist after the lift-off procedure, as
ascertained by the atomic force microscopy, and we do not consider
them to have any impact on the properties of the samples.

The superconducting properties of these structures were
investigated with transport measurements, applying the magnetic
field perpendicularly to the sample surface and using a transport
current with the effective value of $100\,$nA and frequency of
$27.7\,$Hz. The measurements were taken with the field and
temperature resolution of $10\,{\rm \mu T}$ and $0.4\,{\rm mK}$,
whilst the resolution of the DC current was $0.01\,{\rm \mu A}$. A
special attention was being paid to eliminate any possibility of a
trapped flux in the setup before a set of measurements was started
up.

Fig. \ref{icb} shows the critical current of the structures
(filled symbols) versus the normalized applied flux, taken with
the step of $0.1\Phi_{a} /\Phi _{0}$ within the range
$\Phi_{0}\leq \Phi_{a}\leq \Phi_{0}$ at $99.5\,\%$ of the maximum
critical temperature of the structures $T_{cm}=1.3105\,$K. The
solid lines are the theoretical curves obtained by using de Gennes
- Alexander theory for superconducting micronetworks \cite{fink}.
The flux has been calculated with respect to the mean radius
 of the loop $r_{m}=797.5\,$nm, taking the field parallel
to the $z$-direction as positive (Fig. \ref{afm}).

A homogeneous mesoscopic superconducting loop whose width $w$ and
thickness $t$ are much smaller than the coherence length $\xi (T)$
($w,t<<\xi (T)$, 1D-regime), exhibits an oscillatory dependence of
the critical current  \cite{victor,fink}
\begin{equation}
I_{c}=I_{cm}|{\rm cos}(\pi {\Phi \over{\Phi _{0}}})| \label
{squid}
\end{equation}
where $\Phi $ is the total flux through the loop and $I_{cm}$ is
the maximum critical current. Close to the zero-field critical
temperature the influence of self-inductance can be neglected and
the flux through the loop equals the applied flux. As the
coherence length at $T=0.995T_{cm}$ is $\xi (T)\approx 1.9\,{\rm
\mu m}$ and the width of the loops is $w=0.235\,{\rm \mu m}$, the
samples are in the 1D-regime and the de Gennes - Alexander theory
is applicable.  In order to take into account the dots, Eq.
(\ref{squid}) has to be modified by adding the flux generated by
the dots to the applied flux. The stray fields of the dots were
obtained by magnetostatic calculations \cite{ja,martin}. Using the
saturation magnetization of the bulk Co, it has been estimated
that the flux generated by the dot in the sample A is $-0.4\Phi
_{0}$, whereas in the sample B the flux equals $-1.6\Phi _{0}$.
Both values are in a good agreement with the experimental
$I_{c}(\Phi / \Phi _{0})$ data (Fig. \ref{icb}). Close to the zero
applied flux $\Phi_{a}/\Phi _{0}=0$, sample A has a finite
resistance, which is lower than the residual resistance in the
normal state, and its critical current has therefore been taken as
zero. On the other hand, for the same applied flux sample B
remains in the superconducting state and has a finite value of the
critical current. A very good agreement in the periodicity of the
critical current, but similar discrepancies in the amplitude of
the critical current have already been observed in Ref.
\cite{victor}.

The loops have a minimum in the critical current in the vicinity
of zero applied flux and maxima for finite applied fluxes round
$\pm \Phi _{0}/2$. This is a clear evidence that phase shifts are
introduced in the loops. More importantly, given that the
superconducting loops are identical and that the minima and maxima
are attained for different values of the applied flux, it is
evident that the phase shift is governed by the flux of the dot.
Neither of the curves displays exactly $\pi$-shift, but we have
convincingly demonstrated that magnetic dot can efficiently be
used as a phase shifter, as well as that the phase shift can be
controlled by changing the parameters of the dot. The flux of the
dot necessary for $\pi$-shift can be tuned by changing the radius
of the dot or by varying the number of Co/Pd bilayers. By
increasing the number of Co/Pd bilayers the magnetization, as well
as the coercive field of the dot increase, whereas the direction
of the magnetization, remanence and the shape of the hysteresis
loop remain unchanged. Moreover, the dependence of the coercive
field on the number of Co/Pd bilayers makes it possible to ensure
that an externally applied magnetic field, used for instance for
the read-out of the qubit, does not affect the flux generated by
the dot.

Even though electron beam lithography introduces to some extent
manufacturing errors, additional lithographic step with an
accurate alignment, needed to place a magnetic dot at the centre
of the loop, as well as optimization of the radius of the dot in
order to generate flux needed for exact $\pi$-shift do not
constrain applicability of the phase shifter. A magnetic dot as a
phase shifter is intended for superconducting flux qubits that are
also fabricated by electron beam lithography and, therefore,
posses their own variability in dimensions, as well as properties
of the Josephson junctions \cite{qb}. Equally importantly,
superconducting flux qubits are typically loops with the area of
approximately $20-25\,\mu m^{2}$. In this range, dedicated
electron beam lithography tools are very accurate and there is
effectively no discrepancy between the designed and achieved
dimensions of a structure.

\begin{figure}[hbt]
\centering
\includegraphics*[width=8.5cm]{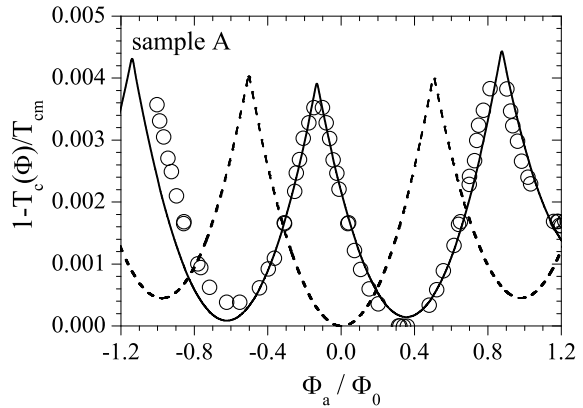}
\caption{The superconducting phase boundary of the sample A
presented as $1-T_{c}(\Phi)/T_{cm}$ versus the $\Phi_{a}/\Phi
_{0}$. The open symbols are experimental data, solid line is the
theoretical fit and dashed line depicts the theoretical phase
boundary of the loop without magnetic dot. \label{tcb}}
\end{figure}

\begin{figure}
\centering
\includegraphics*[width=8.5cm]{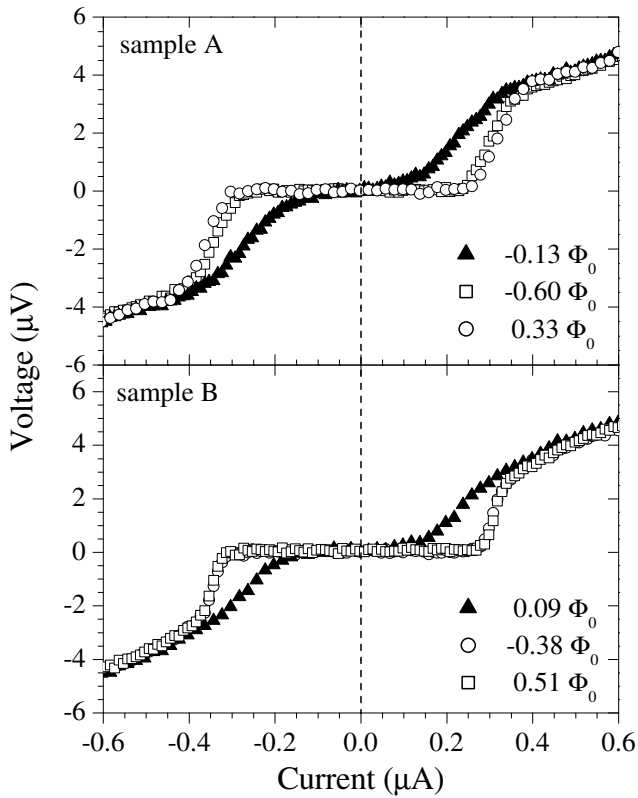}
\caption{I(V) curves of the samples taken at the field values
corresponding to the minima and maxima of the phase boundary
within the range $-\Phi_{0}\leq \Phi_{a} \leq \Phi_{0}.$
\label{ivdc}}
\end{figure}

Fig. \ref{tcb} presents the superconducting phase boundary of the
sample A. The data are given as the normalized critical
temperature $1-T_{c}(\Phi )/T_{cm}$ versus the normalized applied
flux $\Phi_{a} /\Phi _{0}$. The open symbols are experimental
data, solid line is the theoretical curve, whereas the dashed line
is the theoretical curve obtained for the reference loop with the
same parameters, but without magnetic dot. The theoretical
$T_{c}(\Phi /\Phi _{0})$ dependence, found from the
Ginzburg-Landau theory, shows a good agreement with the
experimental data. For the details of our method we refer to Ref.
\cite{japrb}. The real dimensions of the loop, as well as the
aforementioned values of the flux generated by the dots have been
used in the calculations and the best agreement with the
experimental data has been obtained for the coherence length of
$\xi(0)=100\,$nm. The discrepancy between this value of $\xi(0)$
and the estimation from the reference sample $\xi(0)=138\,$nm is
typically encountered in mesoscopic Al structures and is accounted
for by the influence of the contacts, which effectively increase
the radius of the superconducting loop, as well as by minor
nonuniformity in the width of the loop \cite{vicmon}. The
$T_{c}(\Phi )$ phase boundary is pronouncedly modified by the
stray field of magnetic dot, displaying a phase shift. It should
be noted here that, due to its inhomogeneity, the stray field
affects the phase boundary in a nontrivial way bringing about an
additional asymmetry in the phase boundary, as discussed in Ref.
\cite{japrb}. For this reason, two consecutive minima in the phase
boundary of the loop with magnetic dot (corresponding to local
maxima of the critical temperature) have different values and
appear at different fluxes. The inhomogeneity of the stray field,
however, does not prevent the application of the proposed phase
shifter. In order to operate, the superconducting qubit has to be
driven to the state where the flux is $\Phi_{0}/2$. Other states
are less important for the operation of the qubit, provided that
they are well separated from the relevant state so that the
interference between them can be ruled out. Irrespective of its
inhomogeneity, the stray field of the dot can generate
$\Phi_{0}/2$, thus providing the necessary conditions for the
qubit to operate. We mention that in our experiment the
$T_{c}(\Phi)$ phase boundary may be slightly affected by the
displacement of magnetic dot from the centre of the opening, which
is for both samples approximately $125\,$nm and is caused by the
limitations of the electron beam writer at our disposal.

 Fig. \ref{ivdc} presents I(V) curves taken at the
values of the applied flux which correspond to the maxima and
minima of the superconducting $T_{c}(B)$ phase boundary (the phase
boundary of sample B is not shown) at $0.995T_{cm}$. The values of
the fluxes are indicated in the figures. The presence of the phase
shifts is unambiguously and directly demonstrated for both
samples, as the critical currents for finite applied fluxes are
higher than the critical currents
around the zero applied flux. 

In conclusion, we have fabricated and investigated a contactless
magnetic phase shifter for flux-based superconducting qubits. It
has been demonstrated that a phase shift can be induced in the
superconducting circuit, as well as that it can be controlled by
modifying the parameters of magnetic dot. As the magnetic phase
shifter can generate an arbitrary phase shift it can be used as an
external high stability current source for superconducting
elements, or may be deployed as a tool to investigate fractional
Josephson vortices. The experimental results have been interpreted
in the framework of de Gennes - Alexander theory for
superconducting micronetworks and the Ginzburg - Landau theory.

The authors would like to thank M. J. Konstantinovi\'{c} for
valuable discussions and F. Vargas for AFM measurements. This work
has been supported by the Belgian IUAP, the Flemish FWO and the
Research Fund K. U. Leuven GOA/2004/02 programmes, as well as by
the ESF programme "VORTEX". W. V. P. acknowledges the support from
the Research Council of the K.U. Leuven and DWTC. M. M. is a
postdoctoral fellow of IWT-Vlaanderen.

\end{document}